\documentclass[prl,twocolumn,showpacs,preprintnumbers,amsmath,amssymb]{revtex4}

\usepackage{graphicx}
\usepackage{dcolumn}
\usepackage{bm}

\def\lsim{\mathrel{\lower2.5pt\vbox{\lineskip=0pt\baselineskip=0pt
\hbox{$<$}\hbox{$\sim$}}}}
\def\gsim{\mathrel{\lower2.5pt\vbox{\lineskip=0pt\baselineskip=0pt
\hbox{$>$}\hbox{$\sim$}}}}

\newcommand{\im}{{\mbox{Im}\,}}
\newcommand{\re}{{\mbox{Re}\,}}
\newcommand{\be}{\begin{equation}}
\newcommand{\ee}{\end{equation}}

\newcolumntype{L}{>{$}l<{\quad$}}
\newcolumntype{C}{>{$\quad}c<{\;$}}
\newcolumntype{X}{D{\pm}{\pm}{-1}}

\newcommand{\NP}[1]{Nucl.\ Phys.\ {#1}}

\newcommand{\PL}[1]{Phys.\ Lett.\ {#1}}

\newcommand{\PR}[1]{Phys.\ Rev.\ {#1}}

\begin{document}


\title{Chiral condensate thermal evolution at finite
baryon chemical potential \\within Chiral Perturbation Theory}

\author{R. Garc\'{\i}a Mart\'{\i}n and J. R. Pel\'aez }

\affiliation{Departamento de F{\'\i}sica Te{\'o}rica II,
  Universidad Complutense de Madrid, 28040   Madrid,\ \ Spain}


\begin{abstract}
We present a model independent 
study of the chiral condensate evolution in a hadronic gas, in terms
of temperature and baryon chemical potential. 
The meson-meson interactions are described within Chiral Perturbation Theory 
and the pion-nucleon interaction by means of Heavy Baryon Chiral Perturbation Theory,
both at one loop, and nucleon-nucleon interactions 
can be safely neglected within our hadronic gas 
domain of validity. Together with the virial expansion, 
this provides a systematic expansion 
at low temperatures and chemical potentials,
which includes the physical quark masses. This
can serve as a guideline for further studies on the lattice.
We also obtain estimates of the
critical line of temperature and chemical potential 
where the chiral condensate melts, which systematically lie  somewhat higher
than recent lattice calculations but are consistent with several hadronic models.
We have also estimated uncertainties due to chiral parameters, heavier hadrons
and higher orders  through unitarized Chiral Perturbation Theory.
\end{abstract}

\maketitle

\section{Introduction}
One of the most interesting questions about QCD is its
phase diagram, and in particular, the
transition from a hadron gas to a quark gluon
plasma, and the restoration of chiral symmetry
at finite baryon density. On the experimental side,
there are several experiments like CERES, KEK, STAR, SPS, LHC and RHIC
that probe this transition at 
different values of temperature and baryon density.
On the theoretical side, first principle calculations 
on the quark gluon phase performed on the lattice have 
been traditionally hindered by the well known fermion determinant sign problem.
The search for alternative approaches to overcome this problem, at least 
for some values of the chemical potential, 
have recently bolstered the activity in
lattice QCD at finite density 
(see for example \cite{Fodor:2001pe,deForcrand:2002ci,Allton:2003vx,Fodor:2004nz} 
and \cite{Muroya:2003qs} for a review and further references). 
However, lattice studies still present some difficulties in 
the infinite volume extrapolation and most importantly in that they do not use realistic
values for the quark and hadron masses, particularly the pions and kaons,
which are the lightest mesons and the most abundant at low temperatures.

However, from the hadronic phase it is also possible to obtain
model independent predictions by means of Chiral Perturbation Theory (ChPT)
\cite{Weinberg,chpt1,GL3}
which is the low energy Effective Theory of QCD 
(see \cite{books} for introductions and reviews). 
Let us recall that the spontaneous
chiral symmetry breaking of QCD requires the existence of 
eight massless Goldstone Bosons that can be identified with 
the pions, kaons and the eta. Therefore they are the most 
relevant degrees of freedom at low energies. With these fields, 
the ChPT Lagrangian  is built as the most general 
derivative expansion, over $4\pi f_\pi\simeq 1.2\, \hbox{GeV}$
(the symmetry breaking scale), respecting the
symmetry constraints of QCD. Hence ChPT is 
the low energy effective theory of QCD.
Actually, there is also an explicit symmetry breaking
due to the small quark masses that give rise 
to a small mass for the pions, kaons and the eta, which 
are thus just pseudo Goldstone Bosons. For this reason ChPT is 
an expansion also in masses which are treated perturbatively.
The leading term is fixed once
we know the symmetry breaking scale, and the next orders contain
a finite number of constants that absorb the infinities
generated from loops, rendering the calculations
finite order by order. 
Baryons can also be included in the Lagrangian
respecting chiral symmetry but their treatment is more involved
due to their large masses. Within Heavy Baryon 
Chiral Perturbation Theory (HBChPT) \cite{Jenkins:1990jv} 
this problem is overcome
for the meson-baryon interaction
by an additional expansion over the baryon mass.
Since we are only interested in the quark condensate in a
hadronic gas, we 
will neglect the nucleon nucleon interactions. In particular, this leaves
cold nuclear matter outside our domain of applicability.
We therefore have a model independent formalism derived
from QCD that provides a systematic expansion at low energies
and chemical potentials.
This approach has proven very successful and works remarkably well
within the meson sector, whereas for the meson-baryon sector,
the convergence is somewhat slower. 

In order to include the thermodynamic effects of the temperature and chemical
potential, we will use another model independent approach, 
namely, the virial expansion \cite{Dashen,virial2,virial3}.
It is a simple and successful technique already applied to describe
 dilute gases made of interacting pions
\cite{Gerber} and other hadrons \cite{virial3}. 
In contrast to lattice approaches, it is very straightforward to
introduce the baryon chemical potential.
For most thermal observables it is enough to know
the low energy scattering phase shifts of the particles within the gas,
which could be taken from experiment, avoiding
any model dependence.
However,  since we are interested in the 
quark condensate, defined as a derivative of the 
pressure with respect to the quark masses, 
one needs a model independent theoretical description
since it cannot be obtained directly from experiment. 
Thus, in this work we extend to finite baryon chemical potential, $\mu_B$,
the model independent approach that combines the virial expansion and ChPT,  
a method already applied at $\mu_B=0$ in
\cite{Gerber,Dobado:1998tv,Pelaez:2002xf}.

For the condensate, it is therefore particularly important to note that
we are using the physical hadron masses in all our calculations.
Let us also remark that the expansion that relates hadron masses to quark
masses, particularly for pions,
shows some of the best convergence properties
in ChPT, much better than the energy expansion itself.
We therefore hope that our results at low
energy and chemical potential could serve as a guideline for
a correct inclusion of mass effects in further studies, for instance,
on the lattice, at least in the isospin limit.

The plan of the work is as follows. In the next section we introduce
in detail the virial formalism and make a rough estimation, using the free gas,
of its applicability bounds in the temperature-chemical potential plane.
With a  brief introduction to ChPT  and the
hadron mass dependence on quark masses, 
we incorporate the interactions, which we do in two 
different scenarios. A pure SU(2) gas of pions and nucleons, where we
include systematically contributions up to a given order, and 
a more realistic gas including kaons, etas and heavier hadrons, 
where we neglect certain contributions  due to the Boltzmann suppression.
We will show the size of the different terms paying particular
attention to the pion-nucleon interaction, and we will
provide phenomenological parametrizations of the 
condensate melting temperature. In addition we will estimate by extrapolation
the melting line in the temperature-chemical potential plane.
Finally, we will calculate our uncertainties due to
the imperfect knowledge of chiral parameters and 
of the high energy behavior. The latter will be done by
using unitarized chiral amplitudes.
We will conclude with a summary and discussion of our results.

\section{The virial expansion}

The thermodynamics of a system of hadrons 
is encoded in the grand canonical potential density
$z=\epsilon_0-P$, where we have explicitly $\epsilon_0$, which
 is due to the existence of a vacuum expectation value even at $T=0$, 
and $P$ stands for the pressure~\cite{Gerber,kapusta}.
In our case we are interested in a multicomponent 
relativistic gas made of pions, kaons, etas and nucleons.
Later on we will introduce heavier hadrons.
In addition, we will assume that only the strong
interactions are relevant and that the baryon density,
defined as $n_B-n_{\bar B}$ is conserved. 
For that reason,
the pressure will depend on the temperature $T$ and
a baryon chemical potential $\mu_B$. 
Note that the non-strange quark chemical potential $\mu_q= \mu_B/3$
is also frequently used in the literature.
The relativistic virial expansion \cite{Dashen,virial2,virial3} reads
\begin{equation}
\beta P = \sum_i \left( B_i^{(1)}\,\xi_i + B_{i}^{(2)}\,\xi_i^2 +
\sum_{j\ge i} B_{ij}^{int}\,\xi_i\xi_j + \dots \right),
\label{eq:virial}
\end{equation}
where $\xi_i=\exp[\beta(\mu_i-M_i)]$,  $\beta=1/T$, $M_i$
is the mass of the $i$ species and $\mu_i=0,\pm\mu_B$ for mesons,
baryons (antibaryons), respectively.
The coefficients 
\begin{gather}
B_i^{(n)} = \frac{g_i \eta_i^{n+1}}{2\pi^2}\int_0^\infty dp\,p^2\,
e^{-n\beta(\sqrt{p^2+M_i^2}-M_i)}\quad,
\label{virialfree}
\end{gather}
correspond simply to the
virial expansion of the pressure for a free gas
\begin{equation}
\label{exact:pfree}
\beta P_{free} = - \sum_i \frac{g_i \eta_i}{2\pi^2} \int_0^\infty dp\,p^2\,\log\left[
  1 - \eta_i e^{-\beta(\sqrt{p^2+M_i^2}-\mu_i)}\right],
\end{equation}
where $\eta_i=1$ for bosons and $\eta_i=-1$ for fermions.
The interactions appear in the virial expansion
through the $S$-matrix~\cite{Dashen,Gerber}.
For the meson-meson and pion-nucleon interactions relevant for
this work, this can be recast
in terms of the elastic scattering phase shifts, thus we can write:
\begin{equation}
B_{ij}^{int}=\frac{e^{\beta(M_i+M_j)}}{2\,\pi^3}\int_{M_i+M_j}^{\infty} dE\, E^2 K_1(\beta E) 
\Delta^{ij}(E),
\label{2vircoef}
\end{equation}
where $K_1(x)$ is the first modified Bessel function of the second kind
and:
\begin{equation}
\Delta^{ij}=\sum_{I,J,S} (2I+1)(2J+1)\delta^{ij}_{IJS}(E),
\label{Delta}
\end{equation}
$\delta^{ij}_{I,J,S}$ being the $ij\rightarrow ij$
phase shifts (defined so that $\delta=0$ at threshold) 
of the  elastic scattering of a state
$ij$ with well defined isospin, 
total angular momentum and strangeness
$I,J,S$. For the nucleon-nucleon interaction, using the $S$-matrix
representation could be more convenient than the phase shifts, 
but here we are only interested in a sufficiently diluted hadron
gas, so that we can neglect $NN$, and the above formalism is enough for our
purposes.
Thus, in this work we will use the virial expansion
together with ChPT phase shifts, extending previous
works to include a non-vanishing baryon chemical
potential in a hadron gas. We will study the validity of 
this approach in a section below.

Let us now recall that the quark masses appear in the Lagrangian
as $m_q\bar q q$, therefore the non-strange quark condensate 
is given by \cite{Gerber}:
\begin{equation}
\langle\bar{q} q \rangle_{T,\mu_B}=\frac{\partial z}{\partial \hat m}
=\langle 0 \vert\bar{q} q\vert 0 \rangle-  
\frac{\partial P}{\partial \hat m}.
\label{condmq}
\end{equation}
Note that we are working in the isospin limit, using a common
mass $\hat m=(m_u+m_d)/2$ both for the u and d quarks,
and $\langle 0\vert\bar{q} q\vert 0 \rangle \equiv\langle 0\vert\bar{u} u+\bar{d} d\vert 0 \rangle$.
The $\langle0\vert\bar{s} s\vert 0 \rangle$
condensate, which is smaller,
and whose thermal evolution is slower than the non strange one 
\cite{Pelaez:2002xf} could also be studied with similar methods,
but this lies beyond our present scope.

\section{Chiral Perturbation Theory}

Let us then briefly explain our use of Chiral Perturbation Theory  (ChPT)
and fix some notation, starting from the purely mesonic sector. 
Since the ChPT Lagrangian is built as the most general derivative and mass
expansion \cite{Weinberg,chpt1,GL3,books}, 
the meson-meson interaction amplitudes
are obtained as a series in even powers of momenta and masses,
 both denoted generically $p^2$.
The leading order starts at $O(p^2)$ and is universal, only depending
on one scale $F$, which at leading order can be identified
with the pion decay constant $f_\pi$. At next to leading order (NLO)
the amplitudes contain one-loop diagrams with $O(p^2)$ vertices 
plus the tree level contributions from the $O(p^4)$ Lagrangian.
Indeed, all one-loop calculations can be renormalized
in terms of a set of $O(p^4)$ parameters, $L_k(\mu)$ and
 $H_k(\mu)$, $\mu$ being the renormalization scale,
which can be determined from a few experiments, or in the case
of $H_2^r$, from the Resonance Saturation Hypothesis \cite{Jamin},
and used for further predictions at low temperatures.
In Table I we list the values of these parameters
 that we will use for the amplitudes in this work.

Within ChPT, partial waves $t_{IJS}$ are obtained as an expansion in
even powers of momenta and masses. Dropping for simplicity the
$IJS$ indices, we find $t(s)=t_2(s)+t_4(s)+...$, where $t_n(s)= O(p^n)$.
Since we need elastic amplitudes for the
virial expansion, the phase shift $\delta_{IJS}(s)$ is simply the complex phase of
its corresponding partial wave. In principle, the ChPT series is only valid
at low energies compared with $4\pi f_\pi\simeq 1\,$~GeV, although in practice
it is limited to momenta of the order of 200-300~MeV above threshold.
For this reason we cannot apply our expansions
for temperatures beyond that range. We will study the applicability of
our approach with more detail in the next section.

\begin{table}
\begin{ruledtabular}
\begin{tabular}{LCC}
 & $ChPT$ & $IAM$ \\
\hline
L_1^r(M_\rho) &   0.4\pm0.3  &  0.561\pm0.008 \\
L_2^r(M_\rho) &   1.35\pm0.3 &  1.21\pm0.001  \\
L_3           &  -3.5\pm1.1  & -2.79\pm0.02   \\ 
L_4^r(M_\rho) &  -0.3\pm0.5  & -0.36\pm0.02   \\
L_5^r(M_\rho) &   1.4\pm0.5  &  1.4\pm0.02    \\
L_6^r(M_\rho) &  -0.2\pm0.3  &  0.07\pm0.03   \\
L_7           &  -0.4\pm0.2  & -0.44\pm0.003  \\
L_8^r(M_\rho) &   0.9\pm0.3  &  0.78\pm0.02   \\
H_2^r(M_\rho) &  -3.4\pm1.1  & -3.4\pm1.1     \\
\end{tabular}
\end{ruledtabular}
\caption{One loop ChPT low energy constants used for our calculations.
Those in the ChPT column are taken from~\cite{BijnensGasser}
and~\cite{chpt1}, whereas those in the IAM column
are taken from~\cite{GomezNicola:2001as}.
$H_2^r(M_\rho)$ is taken from~\cite{Jamin} for both cases.}
\end{table}

\begin{table}
\begin{ruledtabular}
\begin{tabular}{LCC}
 & $HBChPT$ & $HBChPT + IAM$ \\
\hline
a_1                        &  -2.60\pm0.03  &  -1.36\pm0.02  \\
a_2                        &   1.40\pm0.05  &  0.438\pm0.015 \\
a_3                        &  -1.00\pm0.06  & -0.70\pm0.04 \\
a_5                        &   3.30\pm0.05  & 1.29\pm0.04 \\
\tilde{b}_1 + \tilde{b}_2  &   2.40\pm0.3   & 3.06\pm0.3 \\
\tilde{b}_3                &  -2.8\pm0.6    & -0.41\pm0.27 \\
\tilde{b}_6                &   1.4\pm0.3    & -1.5\pm0.2 \\
b_{16}-b_{15}              &   6.1\pm0.6    & 7.4\pm0.5 \\
b_{19}                     &  -2.4\pm0.4    & -3.7\pm0.2 \\
\end{tabular}
\end{ruledtabular}
\caption{Low energy constants for $ O(p^3)$ HBChPT
used in our calculations, for the non unitarized and unitarized
cases. Taken from~\cite{GomezNicola:2000wk}.}
\end{table}

Baryons have a mass of the order of the chiral expansion scale
$4\pi f_\pi$, but they can also be included as degrees of freedom in a chiral
effective Lagrangian if they are treated as heavy particles 
in a covariant framework called Heavy Baryon Chiral Perturbation Theory 
(HBChPT) \cite{Jenkins:1990jv}. In this case, the
series is organized in terms of order $N=1,2,3...$ 
that contain powers of $p^N/(F^{2l}M_N^{N+1-2l})$, where $l=1,...,(N+1)/2$.
Once again, the divergences at each order are absorbed in
a set of parameters, which in the notation of \cite{Mojzis:1997tu}
are called $a's$ at $O(p^2)$ 
and $b's$ at $O(p^3)$.
We provide in Table II the values we have used,
but we want to remark that these parameters are strongly correlated
and other sets can be found in \cite{Fettes:1998ud}.

The HBChPT framework is specially well suited for our approach,
not only since
$\pi N$ scattering has been calculated to one loop, but also because 
it has been unitarized \cite{GomezNicola:1999pu,GomezNicola:2000wk}
describing the $\Delta(1232)$ resonance correctly,
which will be of use for the last section.
In particular, we will use the third order 
calculation \cite{Mojzis:1997tu,Fettes:1998ud}. 
Note that there is a fourth order calculation
available \cite{Fettes:2000xg},
that has also been unitarized \cite{GomezNicola:1999pu,GomezNicola:2000wk},
but it introduces many more chiral parameters (up to 18 parameters in total)
with very large 
uncertainties, and just an slight improvement over the third order results.
Let us also remark that the second order coefficients, which are called $a_i$ 
in the formalism of \cite{Mojzis:1997tu} that we follow here,
can be translated to  other coefficients often used in the literature,
called $c_i$, which are not dimensionless. 
The values listed in the HBChPT column in Table~II correspond to:
\begin{eqnarray}
\nonumber
c_1 = -1.06\pm0.06 \qquad
c_2 =  3.4\pm0.5 \\
c_3 = -5.74\pm0.15 \qquad
c_4 =  3.7\pm0.2,
\label{ourcs}
\end{eqnarray}
in GeV$^{-1}$ units. These are perfectly compatible with 
the recent values given 
in \cite{Meissner}
\begin{eqnarray}
\nonumber
c_1 = -0.9^{+0.5}_{-0.2} \qquad
c_2 =  3.3\pm0.2\\
c_3 = -4.7^{+1.2}_{-1.0} \qquad
c_4 =  3.5^{+0.5}_{-0.2}.
\label{cs}
\end{eqnarray}
The corresponding values used for Unitarized HBChPT in the last section are:
\begin{eqnarray}
\nonumber
c_1 = -0.75\pm0.02 \qquad
c_2 =  1.4\pm0.5 \\
c_3 = -3.1\pm0.3 \qquad
c_4 =  1.5\pm0.2.
\end{eqnarray}

As the last point concerning ChPT, and
as we advanced in the introduction, the chiral condensate virial
calculation requires the knowledge of the quark mass dependence.
Within ChPT, since we deal with hadrons, we actually 
recast Eq.(\ref{condmq}) as follows \cite{Gerber}
\begin{eqnarray}
\langle\bar{q} q\rangle&=&\langle 0 \vert\bar{q} q\vert 0 \rangle -
\sum_h \frac{\partial M_h}{\partial \hat m}\frac{\partial P}{\partial M_h}\\
&=&\langle 0 \vert\bar{q} q\vert 0 \rangle 
\left(1+\sum_h \frac{c_h}{2 M_h F^2} 
\frac{\partial P}{\partial M_h}\right)
\end{eqnarray}
where, for further convenience we have introduced the constant
$F$, which is the pion decay constant in the chiral limit, and
we have defined the coefficients
\begin{equation}
c_h=- F^2\frac{\partial M_h^2}
{\partial \hat m}\langle 0 
\vert\bar{q} q\vert 0 \rangle ^{-1},
\label{cis}
\end{equation}
that encode the hadron mass dependence on the quark mass.
The $c_\pi$ coefficient in SU(2) ChPT was calculated
in \cite{Gerber}. Here we will use the SU(3) ChPT expressions at $O(M_\pi^4)$
which, for $h=\pi, K, \eta$ 
can be read in \cite{Pelaez:2002xf}. Numerically they amount to
\begin{equation}
c_\pi = 0.9^{+0.2}_{-0.4},\quad
c_K = 0.5^{+0.4}_{-0.7},\quad
c_\eta = 0.4^{+0.5}_{-0.7},
\label{eq:coeffs}
\end{equation}
Using the $O(M_\pi^4)$ calculation of the nucleon mass
within HBChPT \cite{Fettes:1998ud, Becher:1999he} we find
\begin{equation}
c_N = 3.6^{+1.5}_{-1.9},
\label{eq:cNcoeff}
\end{equation}
where we have used the recent values~\cite{Meissner} for the
HBChPT parameters in Eq.(\ref{cs}).
This corresponds to a nucleon sigma term:
 \begin{equation}
   \label{eq:sigmaterm}
   \sigma_{\pi N}=40^{+17}_{-21}\ {\rm MeV}
 \end{equation}
where the uncertainties are estimated assuming  uncorrelated errors
and provide a very conservative range.
Had we used the constants in Eq.(\ref{ourcs}), we would have found
$51^{+13}_{-23}\,$ MeV. We have checked that the difference between these
two central values will amount to roughly 1 MeV or less in our melting
temperatures within the validity region of the approach, that we will study in the next section.

Finally, let us analyze the hierarchy of the different terms.
First of all, we see that the contributions 
to the virial expansion of the pressure
are exponentially suppressed
as $\exp{(-\beta M_h)}$. 
Nevertheless, the $\exp(\beta\mu_B)$ 
factor for the nucleons can overcome the previous
thermal suppression if the chemical potential is
of the same order as their mass.
 These exponentials are inherited 
by the derivatives.
In addition, we observe that
$c_\pi/M_\pi > c_N/M_N >> c_K/M_K > c_\eta/M_\eta $.
For the above two reasons, pions and the interactions of the other
species with pions, are the dominant contributions
for studying the melting of the condensate. Nucleons become 
comparable only when their chemical potential 
is of the order of their mass.
In addition, since nucleon-nucleon 
interactions are much
stronger at low energies than those of pion-pion or pion-nucleon
we will carefully exclude the region with too high nucleon density.

As an illustration, and in order to compare with other results 
in the literature as well as to obtain an estimate 
on the reliability of the virial expansion,
we will first study briefly the case of a free gas.

\section{The free gas and the domain of validity of our approach}

The interest of this case is that we have a closed
expression for the pressure, already given in Eq.~(\ref{virialfree}),
which can be integrated numerically and compared with its 
second order virial expansion, namely, Eq.~(\ref{eq:virial})
with all $B_{ij}^{int}=0$, i.e.,
\begin{equation}
\beta P \simeq \sum_h B_h^{(1)}\,\xi_h + B_h^{(2)}\,\xi_h^2.
\label{eq:virial}
\end{equation}
First of all, in Figure 1 we have plotted
the relative abundances 
of the most relevant species as a function of the temperature,
which we parametrize in terms of their density
\begin{equation}
  \label{eq:density}
  n_h(\mu_B,T)=\frac{g_h}{2\pi^2} \int_0^\infty dp\,
                     \frac{p^2}{e^{\beta(\sqrt{p^2+M_h^2}-\mu_h)}-\eta_h}.
\end{equation}
We see that the lightest particles, the pions, are the main component
of the hadronic gas up to temperatures as high as 200 MeV.
As previously remarked, we can also notice
 that nucleons form a small fraction of the gas 
unless we reach high chemical potentials, of the order of 500 MeV or higher.

\begin{figure}[h]
\includegraphics[scale=1.,angle=-90]{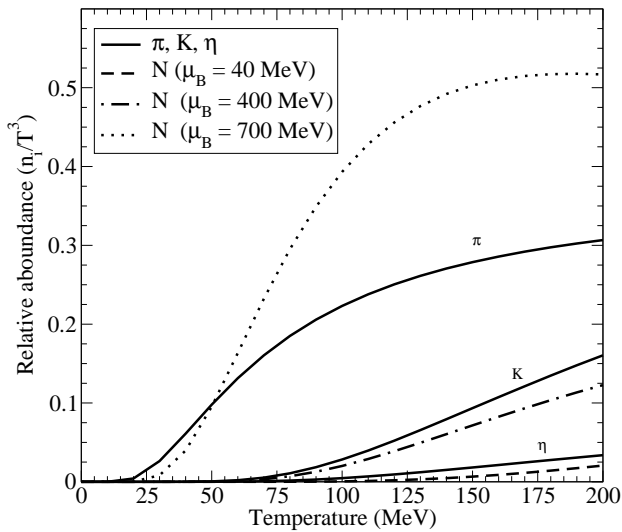}
\caption{\rm Density over $T^3$ of different species in a free hadronic gas.
The pion population is by far the largest up to 200 MeV,
except at very high baryon chemical potentials, where nucleons dominate. }
\end{figure}

Next, in Figure 2 we show, in the $(\mu_B,T)$ plane
the relative error between the exact calculation using Eq.(\ref{virialfree})
and the second order virial expansion 
for the chiral condensate, Eq.(\ref{eq:virial}). This we define as:
\begin{equation}
  \label{eq:relativeerror}
  \epsilon=\frac{
\vert\langle\bar{q}q\rangle_{exact}-\langle\bar{q}q\rangle_{virial}
\vert}
{\frac{1}{2}\vert\langle\bar{q}q\rangle_{exact}+\langle\bar{q}q\rangle_{virial}
\vert}.
\end{equation}
We see that the second order virial expansion 
provides a fairly good approximation $\epsilon\simeq0.05$
for moderate  temperatures  and 
chemical potentials (roughly $T<200$ MeV and $\mu_B<800$ MeV)
and slightly beyond we obtain 
just qualitatively correct results ($\epsilon\simeq0.20$),
since the expansion deteriorates rather rapidly. 
In addition, and since our scope is just to describe a simple hadronic gas,
the nuclear matter regime should be excluded 
from the validity region. In fact, nuclear matter, even at low temperatures,
is not a gas but a fluid \cite{kapusta}, since, at low momenta,
$NN$ interactions are an order of magnitude larger than
$\pi\pi$ or $\pi N$ interactions.
Just for illustration, we have plotted in Fig 2,
as a continuous line,
the points where the saturation density, $\rho_0\simeq0.16\,{\rm fm}^{-3}$, 
is reached. 

Let us however remark that we are not interested in all the hadron
thermodynamics, but just in a particular quantity, the quark condensate,
that does not depend on the interactions themselves 
but on their {\it derivative}
with respect to quark masses.
Thus, the enhancement due to  NN 
interactions with respect to $\pi N$ or $\pi\pi$
is not as large for the quark condensate as it may be 
for other thermodynamic quantities that depend on the $NN$ interactions.
In particular, it has been shown \cite{Lutz:1999vc}, using
an equation of state for the nuclear matter regime 
based on a chiral Lagrangian,
that the evolution of the quark condensate with nuclear density
is well reproduced up to somewhat beyond $\rho_0/2$ just with a linear term
proportional to the nuclear $\pi N$ sigma term, but that
higher orders in density are relevant beyond (see Fig.~3 in \cite{Lutz:1999vc}).
Since the nucleon-nucleon interaction lies outside
the scope of this work, but the $\pi N$ interaction is included,
we only consider densities below $\rho_0/2$. 
Note that with this choice we lie on the safe side. In addition, 
the uncertainties due to the $\pi N$ interaction, that will be studied later
in this work, are much bigger than the neglected
effects of higher orders in density
up to densities much larger than $\rho_0/2$.

\begin{figure}[h]
\includegraphics[scale=1.0,angle=-90]{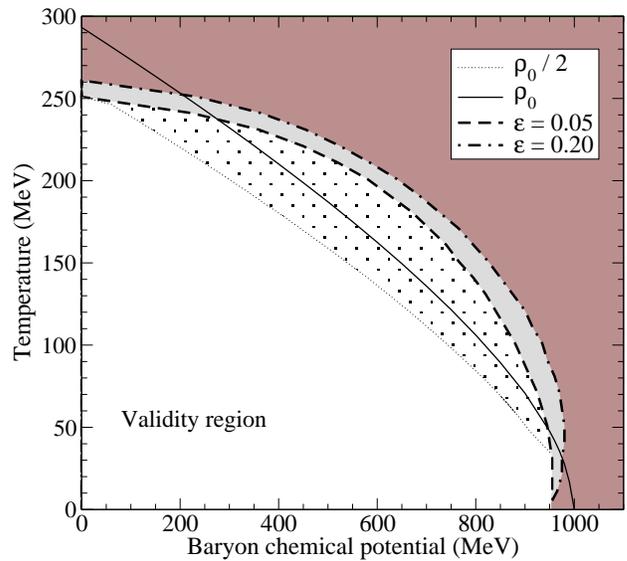}
\caption{\rm Estimation of the region where 
the virial expansion may provide a good approximation to the quark condensate. 
Within the region inside the dashed line, and for a free gas 
of pions and nucleons, 
the relative error $\epsilon$ of the second order
virial expansion with respect to the analytic expression is less than 5 \%.
The virial 
approximation deteriorates rapidly beyond that ``validity region''.
As commented in the text, we also require the nucleon density to be smaller
than the saturation density $\rho_0$ (the thick continuous line) and therefore
we exclude the dotted area. The remaining
white  area is our approximate validity region }
\end{figure}

In summary, in Fig.~2, the dotted area stands 
for densities above $\rho_0/2$ and the remaining white area in Fig.~2
will be referred as the ``validity region'', 
in the understanding that this is just a very crude estimate.
Fortunately, we will see that, once we introduce the
interactions, most of the interesting phenomena, including our
extrapolations for the condensate melting, occur 
within the bounds of this validity region,
so that the virial expansion is reasonably under control
and definitely not diverging wildly. In particular,
the region of relevance 
to study the freeze-out or the phase 
transition in Relativistic Heavy Ion Collisions,
$\mu_B=40 - 50$~MeV, $T\simeq 170 - 200\, $ MeV \cite{Broniowski:2001we} 
lies within our domain of validity.

Finally, we show in Figure 3, the  condensate melting line in
the $(\mu_B,T)$ plane
using either the closed form of the grand canonical potential density or the one
extrapolated from our second order virial expansion. 
First of all, if we only consider $\pi,K, \eta$ and nucleons,
we note that, starting 
from zero chemical potential, the second order virial melting
line 
indeed follows closely the complete calculation with Eq.(\ref{virialfree}),
although it deviates abruptly around $T\simeq180$~MeV and 
$\mu_B=800$~MeV, that nevertheless lies beyond the ``validity region''.

In Figure 3 we also show the melting line that results if we add
the contribution
of the heavier hadrons in the free 
gas approximation by using
\begin{eqnarray}
\Delta \langle \bar{q} q\rangle &=& -\sum_h
\frac{\partial \Delta P}{\partial M_h}
\frac{\partial M_h}{\partial\hat m}\\ \nonumber
&=& \frac{1}{2\pi^2}\sum_h g_h\,M_h
\frac{\partial M_h}{\partial\hat m}\,
\int_0^\infty dp
\,\frac{p^2/\epsilon_h(p)}{e^{\beta(\epsilon_h(p)-\mu_h)}-\eta_h}.
\label{deltaqq}
\end{eqnarray}
with $\epsilon_h(p)=\sqrt{p^2+M^2_h}$.

In such case, the extrapolated melting line
falls slightly outside
the estimated validity region.
We will see in the next sections
that by adding interactions, 
an important part of the melting line moves within the validity region.

Figure 3
is also relevant because it allows us to compare with 
previous results existing in the literature that use the free
hadron gas, Eq.({\ref{exact:pfree}}),
to study the condensate, including also heavier 
particles as above.
In particular we can compare with the melting lines obtained in 
\cite{Nyffeler:1993iz} and \cite{Tawfik:2005qh}. 
We can observe that there is a quantitative difference between those
calculations and the one we present here.
Such difference is mainly due to 
the dependence of the hadron masses on quark masses
$\partial M_h/\partial \hat m$, which in those works was 
not obtained from ChPT, but simply estimated 
as $\partial M_h/\partial \hat m\simeq N_q$ 
(that is the number of light quarks, except for the pion that was
chosen to be $10$)
\cite{Nyffeler:1993iz}  and 
$\partial M_h/\partial M_\pi^2\simeq A/M_h$ with $A\sim 0.9 - 1.2$
\cite{Tawfik:2005qh}.
\begin{table}
\begin{ruledtabular}
\begin{tabular}{LCCC}
c_h    & $ChPT/HBChPT$     & $Nyffeler$ & $Tawfik-Toublan$ \\
\hline
c_\pi  & 0.9^{+0.2}_{-0.4} &     2.0    & 0.8 - 1.1  \\
c_K    & 0.5^{+0.4}_{-0.7} &     0.7    & 1.6 - 2.1  \\
c_\eta & 0.4^{+0.5}_{-0.7} &     0.8    & 2.4 - 3.2  \\
c_N    & 3.6^{+1.5}_{-1.9} &     4.2    & 0.6 - 0.7  \\
\end{tabular}
\end{ruledtabular}
\caption{Coefficients $c_h$ as calculated from estimates in
other hadronic models \cite{Nyffeler:1993iz,Tawfik:2005qh} versus the analytic
values obtained in ChPT or HBChPT.}
\end{table}
Table 3 shows the comparison between the
$c_h$ coefficients obtained from these estimations vs.
the ChPT or HBChPT $O(p^4)$ calculations
in Eqs.(\ref{eq:coeffs}) and Eqs.(\ref{eq:cNcoeff}), which we consider 
very accurate, since the $M_\pi^2/(4\pi f_\pi)^2$ ChPT expansion 
converges very well.
Let us remark that the 
$c_h$ coefficients in those works
are generically larger than ours for $h=\pi$, $K$ and $\eta$.
Hence, the effect of using such estimates is to 
accelerate the thermal melting of the condensate. In particular,
in Figure.3, we show
the results of \cite{Nyffeler:1993iz}, giving rise to a
much lower critical temperature. This explains why
the free gas results in \cite{Nyffeler:1993iz,Tawfik:2005qh}
yield melting temperatures much below those obtained in a free 
gas using ChPT calculations for $\partial M_h/\partial \hat m$,
as in \cite{Gerber,Dobado:1998tv}.
As a matter of fact, we have checked that our approach
reproduces the results in 
\cite{Nyffeler:1993iz} just by changing the $c_h$ coefficients.

In what follows, and given the fact that the free 
and interacting contributions
are separated in the virial expansion,
we show the results both expanding the free gas part to second
order in the virial expansion, and also using the analytic expression.

\begin{figure}[h]
\includegraphics[scale=1.,angle=-90]{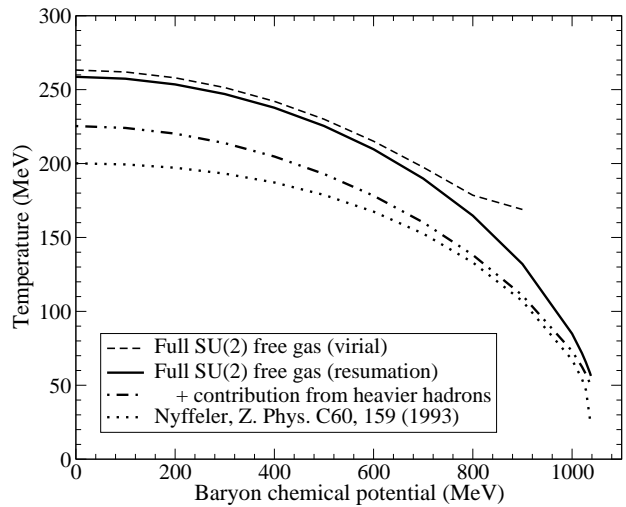}
\caption{\label{fig:epsart} \rm Condensate melting line in the $(\mu_B,T)$
plane for a free pion and nucleon gas (continuous line). Note that it
is very well approximated by the
extrapolated line from the second order virial expansion (dashed)
as long as $\mu_B<800\,$MeV. The melting is faster
when adding free heavier hadrons (dashed dotted line). Finally,
we compare with the free gas results using crude
estimates of $\partial M_h/\partial m_q$ based on the number of 
valence quarks \protect{\cite{Nyffeler:1993iz} (dotted line).}
}
\end{figure}

\section{The interacting SU(2) gas of pions and nucleons}

In the literature it is frequently studied the SU(2) flavor
case \cite{Gerber,Dobado:1998tv,deForcrand:2002ci,Allton:2003vx}, by considering the strange quark as heavy. This is 
simpler, since the pions are the only pseudo-Goldstone bosons 
and there is a clearer suppression of heavier hadrons
due to their heavy masses. On top of that SU(2) ChPT and HBChPT
show a much better convergence than their SU(3) counterparts.
In addition, we can use the one-loop calculations within HBChPT of the
$\pi N$ scattering amplitudes. 

Thus, we will consider the second order virial expansion of a
gas of pions and nucleons, where pions interact among themselves and 
with nucleons. The nucleon-nucleon contribution to the condensate melting 
is suppressed since $c_\pi/M_\pi>c_N/M_N$ and also because
of the Boltzmann suppression of the nucleon population, which can only
be compensated for baryon chemical potentials which lie outside
the validity region  of the virial expansion and are therefore 
beyond our scope.

The $\mu_B=0$ case within ChPT was first studied in \cite{Gerber}
both with the virial expansion and 
by and effective field theory calculation
of the grand canonical potential density. Later on the ChPT virial study was extended 
to a finite {\it pion} chemical potential in \cite{Dobado:1998tv}.
In both works, the only interacting particles were the pions,
and all other hadrons were added in the free approximation.

\begin{figure}[h]
\includegraphics[scale=1.,angle=-90]{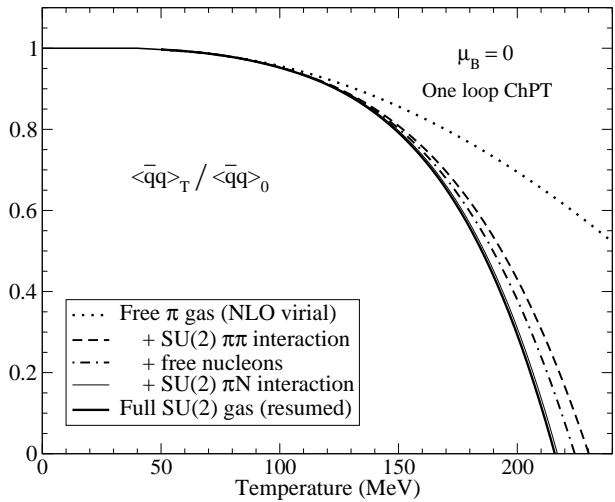}
\caption{\rm Chiral condensate thermal evolution at
zero baryon chemical potential for the SU(2) pion and nucleon gas.
The free gas extrapolated melting temperature falls entirely beyond the 
validity region of the virial expansion, but $\pi\pi$ interactions
bring it below 230 MeV. Although $\mu_B=0$, both 
the free nucleons and their interaction with pions accelerate 
sizably the condensate melting. Note that when
interactions are introduced, the difference between calculating
the free gas contributions exactly (resumed) or with the virial 
expansion is negligible, and the two curves fall on top of each other. }
\end{figure}

In this section, apart from the one loop ChPT $\pi\pi$ scattering
amplitudes, we have included the $O(p^3)$ calculation
of $\pi N$ within HBChPT \cite{Mojzis:1997tu,Fettes:1998ud}. 
First we show in Fig.~4
the results at $\mu_B=0$, and
we can notice how the free SU(2) gas melting temperature
lies beyond the virial validity region, but the pion-pion
interaction brings it down to $\sim 230\,$MeV, within the naive
validity region {\it at $\mu_B=0$}
and in agreement with \cite{Gerber,Dobado:1998tv,Pelaez:2002xf}.
When we further introduce free nucleons 
we get an additional decrease down to 224 MeV.
Of course, within our approach, the melting temperatures
are just extrapolations and lie near the edge of the validity
region.
What is not an extrapolation is the behavior at low T,
where the different contributions are calculated consistently
with the virial and chiral expansions. However, their effect
is more
difficult to see in the figures, and
we quote the extrapolated melting temperatures
 because it is easier to
quantify the relative size of the different contributions.
One should always keep in mind that these are just extrapolations.

Surprisingly, the $\pi N$ interaction, which
is the new contribution that we are adding, {\it has a sizable
effect even at $\mu_B=0$}, decreasing the melting temperature
by another 7 MeV, down to 217 MeV.

Let us emphasize that each additional contribution
decreases further the melting temperature, 
but since the melting of the condensate accelerates near the melting
point (the curve becomes steeper as it gets close to zero),
even if the new contribution has a similar size
as the previous one, its effect on the melting temperature 
seems smaller. That is the reason why the relatively large
effect of $\pi N$ is even more surprising.
A nice illustration of this effect is seen in Fig.~4
since the total result is practically the same
no matter whether we use the virial expansion 
or the analytic (resumed) expression for the free terms.
Indeed, both curves fall on top of each other in the figure.
This is in contrast with Fig.~3, where
there was a difference of roughly 5 MeV observed at $\mu_B=0$
between the free gas melting temperature
depending on whether we used the analytic form of the grand canonical potential density or
its second order virial expansion. The reason is that, in Fig.~3,
after adding all interactions, the slope of the condensate
is so steep that there is almost no difference in the melting
temperature, 1 MeV, between 
using the virial expansion for the free terms or not.

\begin{figure}[h]
\includegraphics[scale=1.,angle=-90]{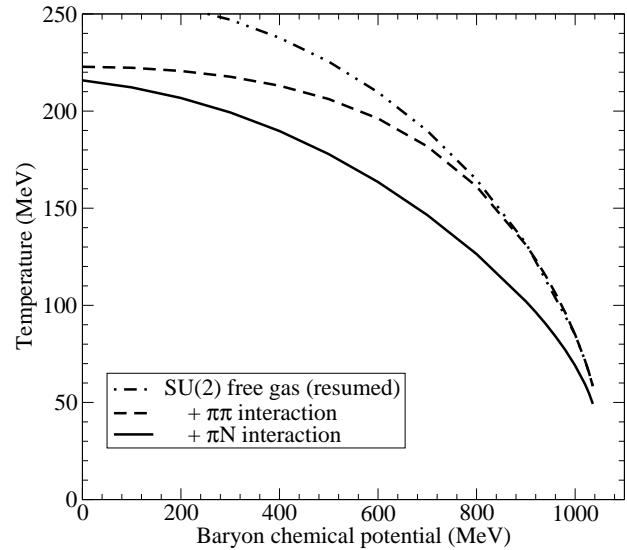}
\caption{\label{fig:epsart} \rm 
Melting line in the $(\mu_B,T)$ plane of the chiral condensate
for an SU(2) gas of hadrons. We show the result for the free gas,
but also adding the $\pi\pi$ ChPT SU(2) interaction to one loop,
and the line resulting from adding the $\pi N$ interaction to third order
in HBChPT. 
}
\end{figure}

Let us now set $\mu_B\neq0$. In Figure 5 we show in the $(\mu_B,T)$ plane
the melting line of the free SU(2) gas of pions and nucleons, the
gas with only interacting pions, and, finally, the effect of adding 
the $\pi N$ interaction. The effect of the latter
becomes bigger as $\mu_B$ increases from zero, and becomes maximum
around $\mu_B=600 - 700$~MeV, where it produces a decrease
in the melting temperature of about 40 MeV with respect to the gas without
$\pi N$ interactions. Up to a chemical potential of $\mu_B=40 - 50$~MeV,
the region of relevance \cite{Broniowski:2001we} 
for Relativistic Heavy Ion Collisions,
the decrease in the melting temperature amounts roughly to
 10 MeV when we include 
the $\pi N$ interaction.

In Fig.~6 we present a comparison of our pure $SU(2)$ results with
lattice estimates in the literature ~\cite{deForcrand:2002ci,Tawfik:2004vv}.
Note that our melting temperature for a given $\mu_B$ lies systematically 
above the corresponding one on the lattice. 
This could already be noticed in the $\mu_B=0$ case \cite{Gerber}, both with
the analytic ChPT calculation of the partition 
function and the virial expansion \cite{Gerber,Dobado:1998tv}. 

\begin{figure}[h]
\includegraphics[scale=1,angle=-90]{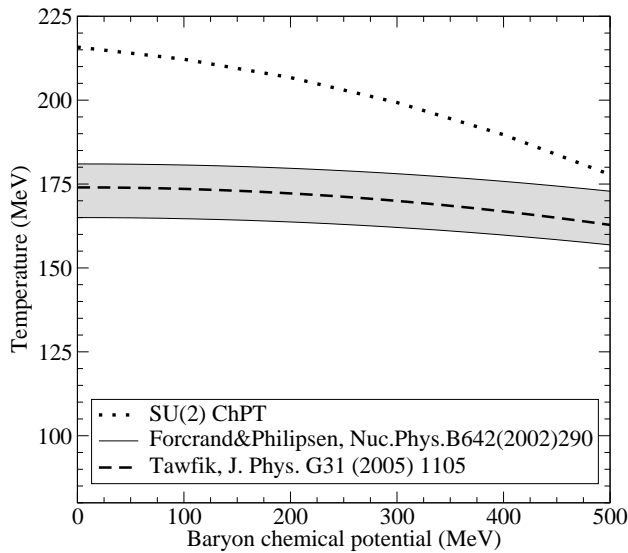}
\caption{Comparison between our results and other estimates in the literature.
The grey band corresponds to the fit in~\cite{deForcrand:2002ci},
while the dashed line corresponds to the one shown in~\cite{Tawfik:2004vv}.
Note that these works obtain values for $T_c$ that are systematically
lower than our model independent approach (dotted line).}
\end{figure}

To facilitate future comparison of our results with other
calculations, we have performed
a simple quadratic fit to our {\it central results} up to $500$~MeV,
giving the following
phenomenological expression for the dependence of the melting
temperature with the baryon chemical potential:
\begin{equation}
 \frac{T_c(\mu_B)}{T_c(0)} = 1 - 0.03155(2)
                      \left( \frac{\mu_B}{T_c(0)} \right)^2.
\label{fitsu2}
\end{equation}
Where the uncertainty is only due to the statistical fit to the
central value, and does not contain the uncertainties 
in our approach, mostly due to the chiral parameters,
 that will be treated in a separated section below.

\section{Realistic hadron gas}

In the previous section we considered a gas just made of pions and 
non-strange nucleons. However, in a real gas, we should consider 
all hadrons. This we will do by including them as free particles.
The only exception will be the kaons and etas, that, 
in view of Fig.~1, are sufficiently
abundant up to 200 MeV to deserve a separate treatment and include
their interaction with a pion. All other interactions are severely suppressed
by Boltzmann and $c_h/M_h$ factors. At very large $\mu_B$ the 
heavier nucleons may not be Boltzmann suppressed,
but that is beyond our estimated validity range. 
 
Thus, in Fig.~7, 
we compare the condensate thermal evolution at $\mu_B=0$
of the SU(2) 
pion and nucleon gas including
$\pi \pi$ and $\pi N$ interactions 
(thick continuous line)
with the results obtained adding further contributions 
that are numerically sizable. 
First, we consider the effect of free kaons and etas,
which accelerate slightly the vanishing of the condensate, decreasing  
the extrapolated melting temperature by $\sim 4\,$MeV. Next, we have 
included the kaon and eta loops in the $\pi\pi$ interaction. This amounts
to calculate $\pi\pi$ scattering in SU(3) ChPT instead of SU(2).
Since the $\pi\pi$ interaction is by far the dominant one, 
this one-loop effect decreases the melting temperature by $\sim 2.5\,$MeV.
When we include the one-loop interactions of $\pi K$ and $\pi \eta$,
we observe a further decrease of $\sim6.5\,$ MeV. 
We have checked that, if we remove the nucleons, these results are in agreement with 
\cite{Pelaez:2002xf}.

To end with the $\mu_B=0$ , we add the heavier hadrons, where we
have estimated that:
\begin{equation}
  \label{eq:heavier}
 \frac{ \partial M_h}{\partial \hat m}\simeq \alpha N_{u,d}^{h}, \qquad \hbox{(heavier hadrons)}
\end{equation}
where $\alpha$ is an adimensional constant
and $N_{u,d}^{h}$ is the number
of valence $u$ or $d$ quarks in the hadron. For nucleons, 
following \cite{Steininger:1998ya}, and from
our Eqs.(\ref{eq:coeffs}) and (\ref{eq:cNcoeff}), we estimate 
$\alpha_N\simeq1.9^{+0.7}_{-1.0}$. For the other heavier
hadrons, we have estimated $\alpha\simeq0.5-2.5$,
which roughly covers the range of $\alpha$ obtained for the 
nucleons and light mesons.

\begin{figure}[h]
\includegraphics[scale=1.,angle=-90]{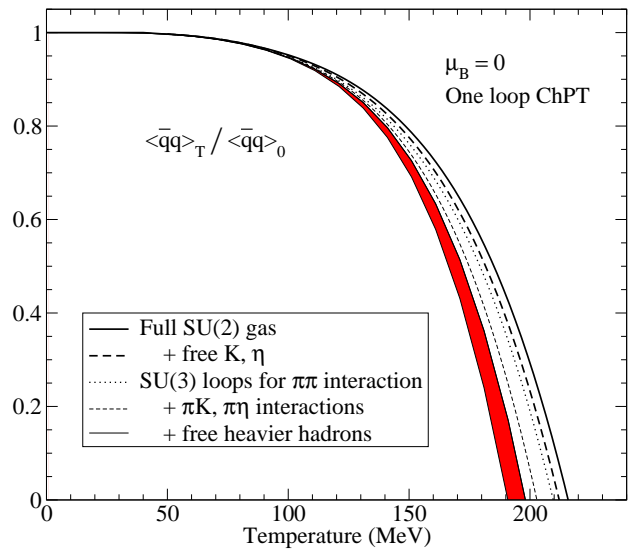}
\caption{\label{fig:epsart} \rm Condensate thermal evolution at $\mu_B=0$
in a gas of hadrons.
Starting from the SU(2) gas including $\pi\pi$ and $\pi N$ interactions,
we show the effect of adding kaons and etas, both as real and virtual states in 
$\pi\pi$ loops, as well as the effect of the $\pi K$ and $\pi\eta$ ChPT one loop
interactions. The band includes the estimated uncertainties due to free heavier hadrons.
}
\end{figure}

Let us now turn to $\mu_B\neq0$ . In Fig.~8 we show
curves for the evolution of the chiral condensate for 
different constant values of the chemical potential.
We can see how the increase of $\mu_B$ accelerates the melting of the condensate,
and that the effect of the chemical potential on the melting
temperature is less than 10 MeV until $\mu_B$ is of order 250 MeV.
In addition, we have plotted the curves as continuous lines within the ``validity region''
estimated in previous sections and as dotted lines outside. In this way we see that
when $\mu_B$ is of the order of the nucleon mass, the aproach deteriorates rapidly, although it
can still give reasonable results at very low temperatures.
A different persrpective is provided in Fig.~9, where we now show the condensate
evolution versus $\mu_B$ for constant $T$. Note that, as we get close to $T=0$ the
evolution with the chemical potential is almost a constant, the condensate
in vaccuum, within
the validity region. At exactly $T=0$ all fugacities vanish, except those of nucleons
when $\mu_B>M_N$, but that lies beyond the validity region and cannot
 be studied with our approach.

\begin{figure}[h]
\includegraphics[scale=1.1,angle=-90]{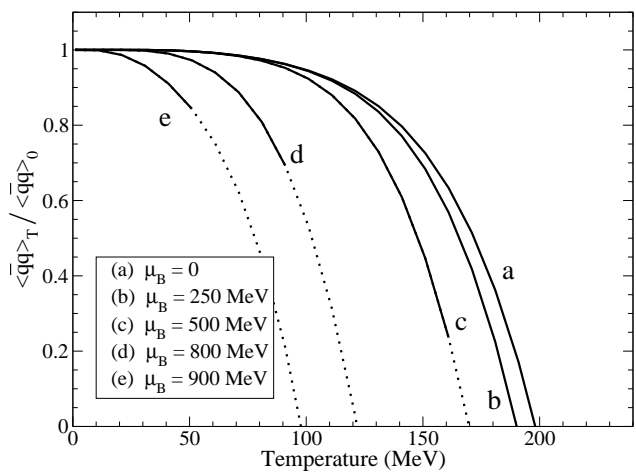}
\caption{\label{fig:epsart} \rm Condensate thermal evolution 
in a gas of hadrons for different constant values of $\mu_B$. The dotted pieces of the curves
lie beyond the ``validity region'' estimated in the text. Note the lines just
correspond to central values.
}
\end{figure}

\begin{figure}[h]
\includegraphics[scale=1.1,angle=-90]{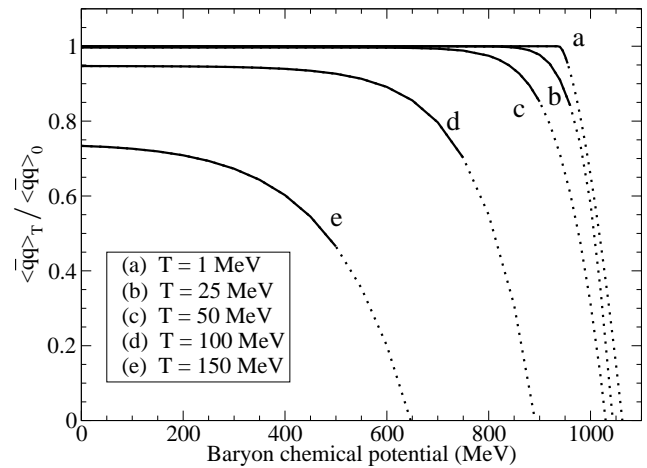}
\caption{\label{fig:epsart} \rm Condensate evolution with the baryon chemical potential 
in a gas of hadrons for different constant values of $T$. The dotted pieces of the curves
lie beyond the ``validity region'' estimated in the text. Note the lines just
correspond to central values.
}
\end{figure}

\begin{figure}[h]
\includegraphics[scale=1.,angle=-90]{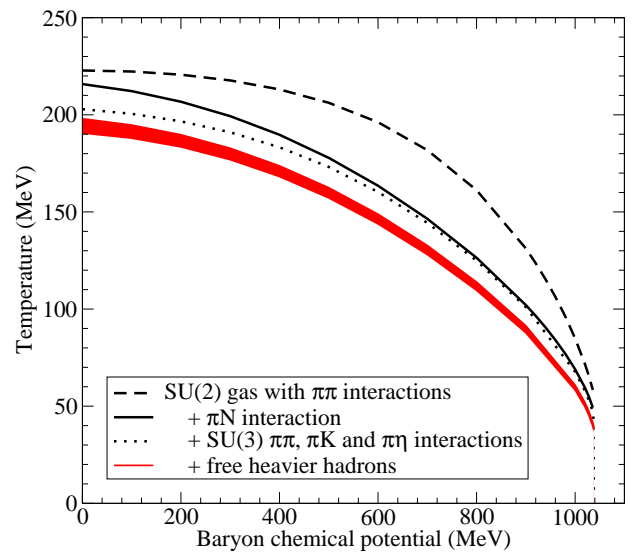}
\caption{\label{fig:epsart} \rm Our final result for 
the chiral condensate melting line in the $(\mu_B,T)$ plane.
Starting from the SU(2) gas with $\pi\pi$ interactions,
we show the effect of $\pi N$ interactions and of adding
kaons and etas. The dark area covers the uncertainty due to 
heavier hadrons estimated as explained in the text.
}
\end{figure}

Finally, in Fig.~10, we present the melting
lines in the $(T,\mu_B)$ plane, showing first the complete SU(2) pion and nucleon gas
result, to which we have added all the effects of kaons and etas.
As it could be expected, the biggest difference is observed at low $\mu_B$.
In particular, in the interesting $\mu_B\simeq40-50\,$ MeV region 
\cite{Broniowski:2001we} we find
that there is a $\sim10$ MeV additional decrease due to kaons and etas,
on top of the one we already found in the previous 
section due to the $\pi N$ interaction.
For higher energies the effect of $K$ and $\eta$ are less evident
not because they decrease, but because they become 
relatively less important with respect to the nucleon free terms and the
due to the larger $\pi N$ interaction.

In the last section we will discuss the comparison of
our results with some previous works in the literature,
but first we will study the uncertainties within our approach.

\section{Uncertainties due to the chiral parameters}

In previous sections we have shown results for the central values
of the chiral parameters given in Table I and II.
In this section we show the uncertainty due to the errors in the 
chiral parameters also listed in those Tables by adding in quadrature
the uncertainties caused by each one of the parameters independently.
Let us remark that the dependence on these parameters is two-fold.
On the one hand, they appear in the interactions through the phase shifts
in Eq.(\ref{Delta}), but, on the other hand, some of them also appear in the
$c_h$ coefficients in Eq.(\ref{cis}) that parametrize the
hadron mass dependence on the quark masses \cite{Pelaez:2002xf}. 
Indeed, the errors in Eqs.(\ref{eq:coeffs})
and (\ref{eq:cNcoeff}) correspond to the uncertainties 
in the chiral parameters.

Thus, in Fig.~11 we show again as a dark band our final extrapolated melting line
in the $(\mu_B,T)$ plane, which includes the uncertainties due to heavier hadrons.
The region between the dashed lines covers also the uncertainties due to the chiral parameters.
The error is highly asymmetric since, as we have already commented, 
it becomes more and more difficult to decrease the melting temperature by adding
new effects, so that the inner dashed line is much closer to our ``central'' results.

\begin{figure}[h]
\includegraphics[scale=1.,angle=-90]{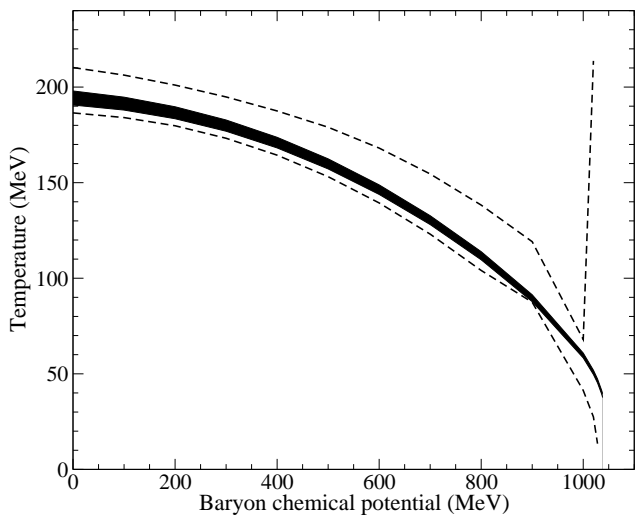}
\caption{\label{fig:epsart} \rm Our final result for
the chiral condensate melting line in the $(\mu_B,T)$ plane.
The dark area covers the uncertainties due to the heavier hadrons, 
that we have included as free components of the gas. The area between
dashed lines corresponds to the uncertainty in the parameters of
ChPT and HBChPT listed in Table I and II.
}
\end{figure}

We also remark that the uncertainties grow very fast at higher $\mu_B$ values
in good agreement with the estimated range of validity of the chiral expansion.
At sufficiently large values of $\mu_B$ our uncertainties are too large
even for qualitative predictions.
However, for small or moderate chemical potentials, say $\mu_B << M_N$
we get quite stable predictions, and the extrapolated melting line is determined
up to a 10 to 15 MeV uncertainty.

\begin{figure}[h]
\includegraphics[scale=1.03,angle=-90]{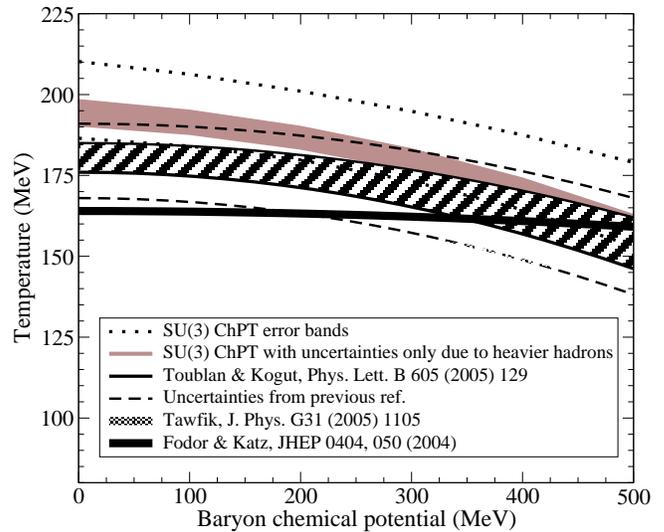}
\caption{Comparison between our results and others in the literature.
Our melting temperatures are consistent within errors with
ref.~\cite{Toublan:2004ks} but systematically higher than \cite{Tawfik:2004vv,Fodor:2004nz}.}
\label{comprealistic}
\end{figure}

In Fig.~\ref{comprealistic} we compare our melting line estimates with previous 
works. We show both our full error band (between the dotted lines)
and that due to heavier hadrons only (light grey band).
Note that we are consistent within uncertainties with other hadronic models
\cite{Toublan:2004ks}, but systematically, our melting temperatures are somewhat 
higher than estimates using lattice \cite{Tawfik:2004vv,Fodor:2004nz}.
Note that, as expected, and in contrast with lattice results, our
melting temperatures always occur few MeV 
above the chemical freeze out temperature
that has been estimated from $p_\perp$ spectra at RHIC
as $T_{f}=165\pm7$~MeV, $\mu_B^f=41\pm5$~MeV~\cite{Broniowski:2001we}.

Finally, to ease future comparisons with our results,
we provide a simple quadratic fit to our results up to $500$~MeV,
giving the following
phenomenological expression for the dependence of the melting
temperature with the baryon chemical potential:
\begin{equation}
 \frac{T_c(\mu_B)}{T_c(0)} = 1 - 0.025^{+0.017}_{-0.011}
                      \left( \frac{\mu_B}{T_c(0)} \right)^2.
\end{equation}
In contrast with the fit given in the pure SU(2) case, Eq.(\ref{fitsu2}),
the errors now are much larger since they 
take into account the uncertainty due to the chiral
parameters and other heavier hadrons.

\section{Unitarized interactions}

Up to this point we have been using a model independent approach.
To that end we have used Chiral Perturbation Theory, which is a low 
momenta expansion, 
to describe the interactions between hadrons. However, in the virial
coefficients, Eq.(\ref{2vircoef}) the momentum integral extends to infinity,
and one may wonder if the extrapolation of ChPT amplitudes to high energies 
beyond their applicability range may be distorting the results. 
In order to check how important is the high energy contribution to those integrals
we will thus extend the ChPT 
amplitudes by means of unitarization, and in particular
the Inverse Amplitude Method (IAM), which provides a remarkably good description
of the meson-meson scattering data up to much higher energies. 
This will introduce a mild model dependency
but will allow us to estimate how large is the effect of the high energy
extrapolation in the virial integrals.

Let us first briefly review unitarization and the IAM. 
For a detailed description of the method and its results 
we refer to \cite{IAM1,IAMcoupled,GomezNicola:2001as,Pelaez:2004xp}.
Note that, in view of Eqs.(\ref{2vircoef}) and (\ref{Delta}),
 we need phase shifts of elastic amplitudes
of definite isospin I and angular momentum J.
Such amplitudes are called partial waves, $t_{IJ}(s)$,
although for brevity we will drop the $I,J$ indices.
Unitarity for elastic partial waves implies that 
\begin{equation}
\im t(s)= \sigma(s) \vert t(s) \vert^2, \qquad \sigma(s)\equiv2 q_{CM}/\sqrt{s},
\label{unit}
\end{equation}
where $q_{CM}$ is the center of mass momentum.
Note that this implies that the modulus of the partial
wave in the elastic regime is bounded by $\left|t(s)\right| < \sigma^{-1}(s)$.

The ChPT partial waves are obtained as an expansion
in momenta and masses $t(s)=t_2(s)+t_4(s)+...$,
where $t_2=O(p^2)$, $t_4=O(p^4)$, where $p$ denotes a generic momenta
or pseudo-Goldstone boson mass. Thus, unitarity
is only satisfied order by order:
\begin{equation}
  \label{eq:pertunit}
\im t_2(s) = 0,\qquad
\im t_4(s) = \sigma(s) \left|t_2(s)\right|^2,\qquad\dots
\end{equation}
This means that,
since the truncated ChPT series grow like polynomials,
unitarity can be badly violated at high energies. 
In addition, typical features of strong interactions,
which are the resonances that saturate the above unitarity bound,
are absent from the ChPT partial waves, since they correspond to poles
in the second Riemann sheet that cannot be described with polynomials.
In practice, pure NLO ChPT only provides a description of meson-meson scattering data
up to energies of the order of 200 MeV above threshold.

However, partial wave unitarity, Eq.(\ref{unit}), fixes completely 
the imaginary part of the {\it inverse amplitude} 
since it follows that $\im t^{-1}=- \sigma(s)$.
Consequently, in the elastic regime a unitary partial wave can be written as
\begin{equation}
  \label{eq:unitamp}
  t(s)=\frac{1}{\re t^{-1}(s) - i\sigma(s)},
\end{equation}
and therefore we only need an approximation to $\re t^{-1}(s)$.
The Inverse Amplitude Method (IAM) simply approximates 
$\re t^{-1}(s)$ by the NLO ChPT result to find:
\begin{equation}
  \label{eq:IAM}
t^{IAM}(s) = \frac{t_2^2}{t_2-\re t_4-i \sigma t_2^2}=\frac{t_2(s)}{1-t_4(s)/t_2(s)},
\end{equation}
where in the last step we have used Eq.(\ref{eq:pertunit}).
In this way we satisfy unitarity while respecting the NLO chiral expansion,
which is recovered at low energies.    In addition, the IAM can generate poles
associated to resonances. The above derivation
has been done for elastic amplitudes and yields amplitudes
that indeed extend the ChPT calculations describing remarkably 
well the $\pi\pi$ and $\pi K$ scattering data
up to roughly the first relevant inelastic threshold \cite{IAM1}.

The formalism can be easily extended to the coupled channel case
\cite{IAMcoupled,GomezNicola:2001as,Pelaez:2004xp}.
Let us illustrate the case when there are two coupled channel 1 and 2,
(that could correspond, for instance, to $\vert1\rangle=\vert(\pi\pi)_{IJ}\rangle$
and  $\vert2\rangle=\vert(\bar{K}K)_{IJ}\rangle$).
We would then have four partial waves for the different 
choices of initial and final states 
$t_{ij}$, with $i,j=1,2$.  Thus,
the unitarity condition has now the following matrix form:
\begin{equation}
\im T = T \Sigma T^*,\quad
T=\begin{pmatrix}t_{11}&t_{12}\\t_{21}&t_{22}\end{pmatrix},\quad
\Sigma=\begin{pmatrix}\sigma_1 & 0 \\ 0 & \sigma_2\end{pmatrix},
\end{equation}
Following a similar argument as for the one channel case above,
we find:
\begin{equation}
T^U = T_2 \left( T_2 - T_4 \right)^{-1} T_2,
\label{IAMcoupled}
\end{equation}
This expression has been shown to provide a very good description of
meson-meson scattering up to $\sqrt{s}\simeq\,$1.2 GeV, 
much beyond the applicability range of standard ChPT. 
Note that this  has been achieved with chiral parameters $L_i$ compatible
with those of ChPT and therefore with a simultaneous description
of the low energy and resonant regions \cite{GomezNicola:2001as,Pelaez:2004xp}. In particular, 
the $f_0(600)$, $\kappa(900)$, $\rho(770)$, $K^*(892)$, 
$f_0(980)$ and $a_0(980)$
masses and widths are well described together with their associated poles in
the second Riemann sheet \cite{Pelaez:2004xp}.

Concerning the pion-nucleon interactions, the unitarization 
could follow a similar line~\cite{GomezNicola:1999pu}, but it can be improved \cite{GomezNicola:2000wk}
taking into account that HBChPT has a double expansion:
the low momenta chiral expansion, that counts powers of $1/f_\pi$
and a heavy fermion expansion in powers
of $1/M_N$, where $M_N$ is the nucleon mass. 
Factorizing explicitly the powers of $M_N$ and $f_\pi$,
the expansion now reads
\begin{equation}
t=
{M_\pi\over f_\pi^2} t^{(1,1)} +
{M_\pi^2 \over f_\pi^2 M_N } t^{(1,2)} +
{M_\pi^3 \over f_\pi^4 } t^{(3,3)}
+ {M_\pi^3 \over f_\pi^2 M_N^2 } t^{(1,3)} +...,
\label{eq:fexpansion2}
\end{equation}
$t^{(n,m)}$ being dimensionless functions of $\omega/M_\pi$,
where $\omega$ is the pion energy.
Note that, contrary to the meson-meson  case,
odd orders of momenta do occur in the expansion.

The partial wave
unitarity condition with one fermion in the final state simply reads: 
$\im t= q_{CM} \vert t\vert ^2$, so that now
\begin{equation}
  \label{eq:unitamp}
  t(s)=\frac{1}{\re t^{-1}(s) - i q_{CM}}.
\end{equation}
Once again \cite{GomezNicola:1999pu} we can use the HBChPT series at NLO
to approximate $\re t^{-1}(s)$. However, the HBChPT series converges
much worse than in the meson-meson sector, and for that reason
it is convenient to  reorder the heavy baryon expansion (see 
ref.\cite{GomezNicola:2000wk}), 
to get the following unitarized
amplitude:
\begin{eqnarray}
&&t^{IAM}\simeq
\label{eq:UHBChPT}\\
&&\frac{1}{
\frac{f_\pi^2}{M_\pi}\left[
  t^{(1,1)}+\frac{M_\pi}{M_N}t^{(1,2)}+\frac{M_\pi^2}{M_N^2}t^{(1,3)}
\right]^{-1}-M_\pi\frac{t^{(3,3)}}{\left(t^{(1,1)}\right)^2}
}.
\nonumber
\end{eqnarray}
It has been shown that this expression greatly improves the plain
HBChPT description, extending it at least to the nearest relevant
inelastic threshold, and in particular up to $\sqrt{s}\simeq 1400\,$MeV
in the $\Delta(1232)$ resonance channel, which is the most 
relevant feature of $\pi N$ scattering at low
energies. Both the mass and width of this
 resonance are well described by the unitarized amplitude,
together with its associated pole in the second Riemann sheet. 

We will now use Eqs.(\ref{IAMcoupled}) for meson-meson 
and (\ref{eq:UHBChPT}) for $\pi N$ scattering, in order 
to extend ChPT at high energies and check the contribution
of the high momenta region in the virial integrals.
The chiral constants we used for the unitarized 
amplitudes are also listed in Tables I and II. Note that 
for meson-meson scattering they are compatible with the 
values used in standard ChPT at NLO.
The HBChPT parameters are not as well determined as those
in the meson-meson sector, but it is important to
note that the unitarization results are obtained with
chiral parameters of natural size and quite
compatible with different determinations from plain HBChPT.

At this point we want to remark that since we are
generating heavier resonances  
when using the unitarized amplitudes, 
we do not have to introduce them again as free particles
in the gas, as we did in the previous section, to avoid double counting.
In particular, we do not add now the $f_0(600)$, 
$\kappa(900)$, $\rho(770)$, $K^*(892)$, $f_0(980)$, $a_0(980)$
and $\Delta(1232)$ resonances as free particles using Eq.(\ref{deltaqq}),
since they are generated by the IAM.

One advantage of unitarization is that some of the resonances 
are now included with their actual width 
(that is, interacting with the mesons),
which is an effect that we could not take into account by including
them as free components of the gas. A priori, this could be relevant,
since these resonances have strong decays, i.e. large widths, 
and considering them
stable is one of the greatest over-simplifications in hadronic models. We are thus quantifying 
also that effect in our previous model independent parametrizations.
Furthermore, we had a large uncertainty in their $c_h$ or $\alpha_h$ 
parameters,
which is now largely reduced since they appear in the meson-meson
or meson-nucleon virial terms, whose $c_h$ coefficients, Eqs.~(\ref{eq:coeffs})
and (\ref{eq:cNcoeff}),
are much better determined.

Thus, in Fig.~13, we compare the condensate thermal evolution at $\mu_B=0$
that we obtained
in previous sections 
with the results using the unitarized interactions.
The dark narrow band covers, for the IAM results,
the uncertainty due to heavier hadrons
for the non-unitarized results.
As commented in the previous paragraph,
the fact that unitarized interactions include the effect of the 
first heavier resonances, 
accelerates the melting.
In addition, since in the unitarized case 
what we call ``heavier hadrons'' are now the rest of hadrons, which are
heavier and even less abundant, their associated
 uncertainty is much smaller. That is the reason why the dark band from the IAM
is much narrower than the
light band. Let us however, note that, once the uncertainties
are included, the unitarized and non-unitarized
results are very close, showing a remarkable stability of our results
at $\mu_B=0$.

\begin{figure}[h]
\includegraphics[scale=1.,angle=-90]{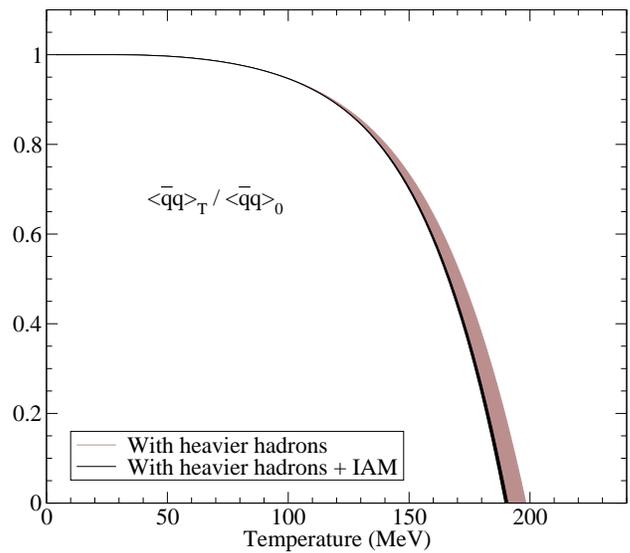}
\caption{\label{fig:epsart} \rm 
Condensate thermal evolution at $\mu_B=0$ with ChPT unitarized
interactions. The resulting melting temperature is slightly lower
but very close to the model independent non-unitarized case
within uncertainties.
}
\end{figure}

Similar considerations hold for the uncertainty bands in Fig.~14.
Again, the light band stands for the non unitarized standard ChPT,
and the much narrower dark band covers the uncertainty due to heavier hadrons
of the IAM result. In this case, we see that the effect of unitarizing is larger,
and the melting is systematically accelerated at moderate $\mu_B\neq0$.
The bigger difference is now due to the fact that, as pointed out above, 
the HBChPT series converges much
worse than the meson-meson one, and the effect of unitarization is 
more dramatic, particularly for the $\Delta(1232)$ that lies very close to threshold.
Still these results give support to 
our statement that our melting temperature estimates 
should be considered as upper bounds, although the differences between
unitarized and non-unitarized are of about 10 MeV at most.

\begin{figure}[h]
\includegraphics[scale=1.,angle=-90]{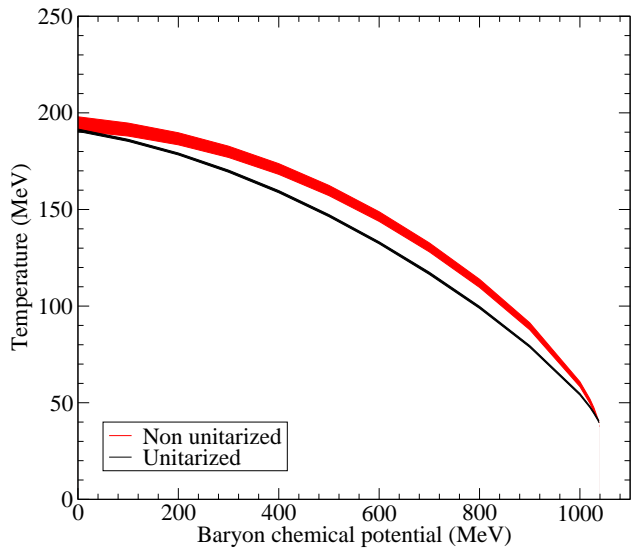}
\caption{\label{fig:epsart} 
Condensate melting lines in the $(\mu_B,T)$ plane with unitarized 
ChPT interactions compared to the non-unitarized ChPT.
The unitarized case melts faster than the non-unitarized, due to
the dynamically generated resonances, that have their correct widths.
}
\end{figure}

Figure~15 shows the errors over this last band due to the uncertainties
in the chiral parameters. The error is smaller than for the non-unitarized
case. It should also be noted that both unitarized and non-unitarized
cases overlap within their errors up to $\mu_B\simeq 300$~MeV and beyond
that are never further than two standard deviations away.

\begin{figure}[h]
\includegraphics[scale=1.,angle=-90]{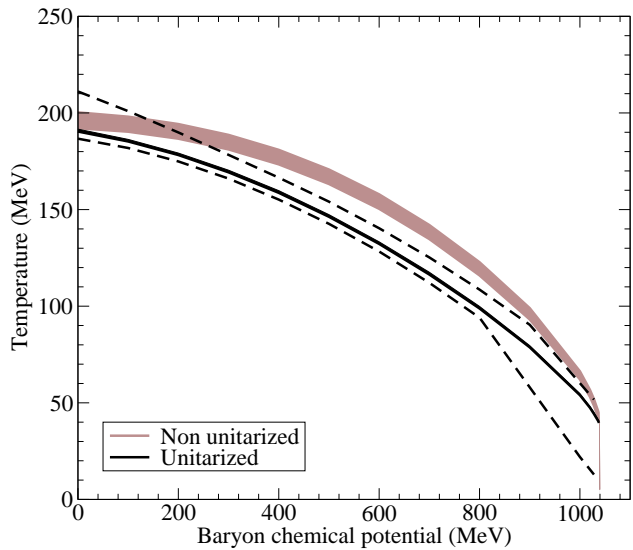}
\caption{\label{fig:epsart} 
The chiral condensate melting line in the $(\mu_B,T)$ plane
for unitarized ChPT interactions. Note that the 
uncertainties due to the heavier hadrons (dark area) are much
smaller than for the non-unitarized case in Fig.~11. The area between
dashed lines corresponds to the uncertainty in the parameters of
unitarized ChPT and HBChPT listed in Table I and II, and is also
smaller than for the non-unitarized case. 
}
\end{figure}

Once again, the following 
quadratic fit 
provides a fairly good representation of our unitarized
results up to $500$~MeV
\begin{equation}
 \frac{T_c(\mu_B)}{T_c(0)} = 1 - 0.032^{+0.011}_{-0.001}
                      \left( \frac{\mu_B}{T_c(0)} \right)^2.
\end{equation}
Where, once more, the errors take into account the uncertainty due to both the chiral
parameters and the heavier hadrons

\section{Summary and discussion}

In this work we have presented a model independent description of a hadronic gas,
based on the virial expansion and Chiral Perturbation Theory (ChPT) to NLO for the 
meson-meson interactions and third order for pion-nucleon within 
Heavy Baryon Chiral Perturbation Theory (HBChPT). 
In particular, we have studied the melting of the
chiral condensate $\langle\bar{q}q\rangle$ as a
function of temperature, $T$, and
baryon chemical potential, $\mu_B$.
We have studied first the validity region, where 
the second order virial expansion is expected to
provide a reasonably good description
of the condensate evolution and nucleon density is low
enough so that we can neglect $NN$ interaction. Since the virial
expansion is a low density expansion
and ChPT is a low energy effective theory,
our approach is best applicable at low
temperature and chemical potential,
although it seems to provide still a fairly
good description of the system up to temperatures
of the order over $T=200$~MeV at $\mu_B=0$
and up to roughly $\mu_B\leq900$~MeV at very low $T$ (see Fig.~2).
This allows us to quantify 
the size of different
contributions to the evolution of the condensate in terms of its extrapolated melting
temperature and chemical potential. Very likely this is a good estimate of 
the transition from ``normal'' hadronic matter, where chiral symmetry is 
spontaneously broken, to a phase where a chiral symmetry restoration may occur.  

We use the virial expansion because it allows us to include easily both 
the chemical potential and the interactions
in the thermodynamics of the system. The chemical potential is very hard
to implement on the lattice, to study rigorously QCD, and our results
may give some guidance on this respect. The interactions 
are included from the zero temperature and density elastic 
phase shifts, which are fairly well known from experiment.
As a matter of fact, for most thermodynamical quantities,
one could simply take the phase shifts from data. However, the
calculation of the condensate requires the knowledge of the
dependence on the quark masses, which is not given by experiment.
Hence, a theory is needed to describe those data, and since perturbative QCD itself 
does not describe the low energy regime, one must turn to ChPT,
which is the low energy effective theory of QCD,
for a model independent description.

Our use of ChPT is very relevant since
it provides model independent and very reliable calculations of the hadrons mass
dependence on quark masses, which is a critical quantity to 
follow the condensate evolution.
This dependence has been frequently estimated roughly in the literature,
but the ChPT calculation provides an expansion in terms of meson masses
which shows a good convergence for pions, kaons, etas and nucleons, 
which are the dominant components of the hadron gas.
Furthermore, we have shown that several results obtained from free hadron gases in 
the literature can be reconciled with the interacting gas predictions if one
uses these correct dependencies.

Next, we have presented results for an SU(2) gas including only pions and nucleons.
This is of interest because its simplicity and because we can include systematically
all the effects up to a given order. 
This is in contrast with what we have called a realistic gas, where we have
considered the interactions of kaons and etas with pions within the
SU(3) ChPT formalism; although, formally,
 the KK, $\eta\eta$ and $\eta K$ interactions
are of the same order, we have neglected them due to: first, the thermal suppression
of kaons and etas compared with the pions and, second, the weaker
dependence of their mass with respect to non-strange quark mass.

In this work we have shown explicitly
the size of different contributions to the condensate melting for $\mu_B\neq0$.
In particular, we have evaluated the free kaon, eta and nucleon contributions,
kaon and eta virtual effects in $\pi\pi$ interactions, as well as
the $\pi K$, $\pi \eta$ interactions, these have also been calculated
at $\mu_B=0$ without nucleons to check with previous results in the literature.
In addition, we have included the $\pi N$ interaction, showing that it does
have a sizable effect {\it even at  $\mu_B=0$}. Around
$\mu_B=40-50\,$MeV the pion nucleon interaction itself
decreases the extrapolated melting temperature by 10 MeV.
At higher baryon chemical potentials, the effect is larger.

It is also relevant to remark that all contributions {\it decrease}
the extrapolated melting temperatures. Intuitively, adding more states
increases the possibilities of generating disorder, i.e., entropy,
and accelerates the melting of the condensate. The interactions seem to 
reinforce this effect. Therefore, our extrapolated melting temperatures can be
considered at least as upper bounds. Nevertheless, we have shown that it becomes harder and harder to decrease the melting temperatures 
with new contributions, so that we still consider that, within uncertainties,
our results provide very good estimates of the melting temperatures.

In this work we have also performed a detailed study of uncertainties, 
which are predominantly of two kinds: on the one hand we have approximated
the effect of heavier hadrons by adding them to the gas as free components. 
The biggest uncertainty comes from the estimated dependence of 
their masses on the non-strange quark mass. 
On the other hand, we have large uncertainties from the chiral parameters
of ChPT and HBChPT. This is the dominant source of error.
All in all, our extrapolated melting temperatures come out with an error of the order 
of $\pm10\,$MeV, somewhat smaller at $\mu_B=0$ and somewhat larger as $\mu_B$ increases.

Let us remark that our results are consistent with other $\mu_B=0$
model independent results that use ChPT and calculate
the partition function either analytically \cite{Gerber} or
also with the virial expansion~\cite{Gerber,Dobado:1998tv,Pelaez:2002xf}.
We also find agreement within errors with  hadronic models \cite{Toublan:2004ks}, but 
our melting temperatures come out systematically higher than 
estimates using lattice~\cite{Tawfik:2004vv,Fodor:2004nz}.
In addition, our melting curves lie, as expected, above the
chemical freeze out temperature that has been estimated from
$p_\perp$ spectra at RHIC~\cite{Broniowski:2001we}.

Finally, in order to have a more realistic description of the high momentum
part of the virial integrals,
we have used a unitarized version of ChPT that extends its applicability
up to $\sim 1.2$~GeV, and of HBChPT that generates the $\Delta(1232)$ resonance. 
It also allows to describe correctly the decay widths 
of the lightest resonances, which are the most abundant of the ``heavier hadrons''
that we had approximated as free components. We have found that our results
are rather stable, within uncertainties, to this changes at $\mu_B=0$, but that
these effects could produce a further decrease of the melting temperatures 
of $\sim 10$~MeV for higher chemical potentials.

In conclusion, we have presented a model independent and systematic
study of the chiral condensate evolution with temperature and chemical potential,
by means of the virial expansion and Chiral Perturbation Theory.
The highlights of this approach with respect to
other fundamental approaches like QCD on the lattice are: 
the use of physical masses for the hadrons,
the good control of the hadron mass
dependence on quark masses, and the simple 
implementation of the baryon chemical potential.
We hope that our results could serve as a guideline for future 
works in hadronic models and also on the lattice, in order
to understand the phase diagram of QCD.

\section{Note added}

After submitting to publication this work, we
became aware of a recent lattice article \cite{Cheng:2006qk},
contested in \cite{Aoki:2006br}, 
where they obtain at $\mu_B=0$, a $T_c=192(7)(4)\,$MeV, 
somewhat higher
than all other lattice results 
we are aware of, and in fairly good agreement
with the ChPT results \cite{Gerber,Dobado:1998tv,Pelaez:2002xf} and here.

\section{Acknowledgments}
Research partially funded by Spanish CICYT contracts
FPA2005-02327, as well as Banco Santander/Complutense
contract PR27/05-13955-BSCH. J.R.P. research is
also partially funded by  CICYT contract BFM2003-00856 
and is part of the EU integrated
infrastructure initiative HADRONPHYSICS PROJECT,
under contract RII3-CT-2004-506078.
R.G.M research is supported by a Doctoral Complutense fellowship.


\begin{thebibliography}{99}
\footnotesize

\bibitem{Fodor:2001pe}
  E.~B.~Gregory, S.~H.~Guo, H.~Kroger and X.~Q.~Luo,
  Phys.\ Rev.\ D {\bf 62}, 054508 (2000)
  Z.~Fodor and S.~D.~Katz,
  JHEP {\bf 0203}, 014 (2002)
  C.~R.~Allton {\it et al.},
  Phys.\ Rev.\ D {\bf 66}, 074507 (2002)
  P.~de Forcrand and O.~Philipsen,
  Nucl.\ Phys.\ B {\bf 673}, 170 (2003)



\bibitem{deForcrand:2002ci}
  P.~de Forcrand and O.~Philipsen,
  Nucl.\ Phys.\ B {\bf 642}, 290 (2002)

\bibitem{Allton:2003vx}
  C.~R.~Allton, {\it et al.}
  Phys.\ Rev.\ D {\bf 68}, 014507 (2003)

\bibitem{Fodor:2004nz}
  Z.~Fodor and S.~D.~Katz,
  JHEP {\bf 0404}, 050 (2004)

\bibitem{Muroya:2003qs}
  S.~Muroya, A.~Nakamura, C.~Nonaka and T.~Takaishi,
  Prog.\ Theor.\ Phys.\  {\bf 110}, 615 (2003)

\bibitem{Weinberg} S. Weinberg, Physica A96, (1979) 327.
\bibitem{chpt1} J. Gasser and H. Leutwyler, Ann. Phys. 158, (1984)
142. 
\bibitem{GL3}J. Gasser and H. Leutwyler, Nucl. Phys. B250,
(1985) 465,517,539.
\bibitem{books}
A. Dobado, A.G\'omez Nicola, A. L. Maroto and J. R. Pel\'aez,
{\it Effective Lagrangians for the Standard Model},
Texts and Monographs in Physics. ed: Springer Verlag,
 Berlin Heidelberg New.York (1997).
A. Pich, Rept.Prog.Phys.58 (1995),563 610.
U.G. Mei{\ss}ner, Rept.Prog.Phys.56 (1993),903 996.

\bibitem{Jenkins:1990jv}
  E.~Jenkins and A.~V.~Manohar,
  Phys.\ Lett.\ B {\bf 255}, 558 (1991).
  V.~Bernard, N.~Kaiser and U.~G.~Meissner,
  Int.\ J.\ Mod.\ Phys.\ E {\bf 4}, 193 (1995)

\bibitem{virial2} G.M. Welke, R. Venugopalan and M. Prakash,
\PL{B245}(1990) 137.
V.L. Eletsky, J. I. Kapusta and R. Venugopalan,
\PR{D48} (1993)4398.

\bibitem{virial3} R. Venugopalan, M. Prakash, \NP{A546} (1992)718.

\bibitem{Dashen} R. Dashen, S. Ma, H.J. Bernstein,\PR{187} (1969) 187.

\bibitem{Gerber} P. Gerber and H. Leutwyler, \NP{B321} (1989) 387.

\bibitem{Dobado:1998tv}
  A.~Dobado and J.~R.~Pelaez,
  Phys.\ Rev.\ D {\bf 59}, 034004 (1999)

\bibitem{Pelaez:2002xf}
  J.~R.~Pelaez,
  Phys.\ Rev.\ D {\bf 66}, 096007 (2002)

\bibitem{BijnensGasser} J. Bijnens, G. Colangelo and J. Gasser,
\NP{B427}, (1994) 427.

\bibitem{GomezNicola:2001as}
  A.~Gomez Nicola and J.~R.~Pelaez,
  Phys.\ Rev.\ D {\bf 65}, 054009 (2002)

\bibitem{Jamin} M. Jamin, \PL{B538} (2002) 71.

\bibitem{GomezNicola:2000wk}
  A.~Gomez Nicola, J.~Nieves, J.~R.~Pelaez and E.~Ruiz Arriola,
  Phys.\ Lett.\ B {\bf 486}, 77 (2000)
  Phys.\ Rev.\ D {\bf 69}, 076007 (2004)

\bibitem{Mojzis:1997tu}
  M.~Mojzis,
  Eur.\ Phys.\ J.\ C {\bf 2}, 181 (1998)
  G.~Ecker and M.~Mojzis,
  Phys.\ Lett.\ B {\bf 365}, 312 (1996)


\bibitem{Fettes:1998ud}
  N.~Fettes, U.~G.~Meissner and S.~Steininger,
  Nucl.\ Phys.\ A {\bf 640}, 199 (1998)

\bibitem{GomezNicola:1999pu}
  A.~Gomez Nicola and J.~R.~Pelaez,
  Phys.\ Rev.\ D {\bf 62}, 017502 (2000)

\bibitem{Fettes:2000xg}
  N.~Fettes and U.~G.~Meissner,
  Nucl.\ Phys.\ A {\bf 676}, 311 (2000)

\bibitem{Becher:1999he}
  T.~Becher and H.~Leutwyler,
  Eur.\ Phys.\ J.\ C {\bf 9} (1999) 643

\bibitem{Meissner}
  U.~G.~Meissner,
  PoS {\bf LAT2005} (2006) 009.
  [arXiv:hep-lat/0509029].

\bibitem{kapusta} J.I.Kapusta, {\it Finite-temperature field
theory} (Cambridge University Press, 1989).

\bibitem{Lutz:1999vc}
  M.~Lutz, B.~Friman and C.~Appel,
  Phys.\ Lett.\ B {\bf 474}, 7 (2000)
  [arXiv:nucl-th/9907078].

\bibitem{Nyffeler:1993iz}
  A.~Nyffeler,
  Z.\ Phys.\ C {\bf 60}, 159 (1993).

\bibitem{Tawfik:2005qh}
  A.~Tawfik and D.~Toublan,
  Phys.\ Lett.\ B {\bf 623}, 48 (2005)

\bibitem{Broniowski:2001we}
  W.~Broniowski and W.~Florkowski,
  Phys.\ Rev.\ Lett.\  {\bf 87}, 272302 (2001)
  Acta Phys.\ Polon.\ B {\bf 33}, 1935 (2002)

\bibitem{Tawfik:2004vv}
  A.~Tawfik,
  J.\ Phys.\ G {\bf 31} (2005) S1105

\bibitem{Steininger:1998ya}
  S.~Steininger, U.~G.~Meissner and N.~Fettes,
  JHEP {\bf 9809}, 008 (1998)

\bibitem{Toublan:2004ks}
  D.~Toublan and J.~B.~Kogut,
  Phys.\ Lett.\ B {\bf 605} (2005) 129

\bibitem{IAM1}
  T.~N.~Truong,
  Phys.\ Rev.\ Lett.\  {\bf 61}, 2526 (1988).
  A.~Dobado, M.~J.~Herrero and T.~N.~Truong,
  Phys.\ Lett.\ B {\bf 235}, 134 (1990).
   A.~Dobado and J.~R.~Pelaez,
   Phys.\ Rev.\ D {\bf 47}, 4883 (1993)
  Phys.\ Rev.\ D {\bf 56}, 3057 (1997)


\bibitem{IAMcoupled}
  J.~A.~Oller, E.~Oset and J.~R.~Pelaez,
  Phys.\ Rev.\ Lett.\  {\bf 80}, 3452 (1998)
  Phys.\ Rev.\ D {\bf 59}, 074001 (1999)
  [Erratum-ibid.\ D {\bf 60}, 099906 (1999)]
  F.~Guerrero and J.~A.~Oller,
  Nucl.\ Phys.\ B {\bf 537}, 459 (1999)
  [Erratum-ibid.\ B {\bf 602}, 641 (2001)]

\bibitem{Pelaez:2004xp}
  J.~R.~Pelaez,
  Mod.\ Phys.\ Lett.\ A {\bf 19}, 2879 (2004)

\bibitem{Cheng:2006qk}
  M.~Cheng {\it et al.},
  Phys.\ Rev.\ D {\bf 74}, 054507 (2006)
  [arXiv:hep-lat/0608013].

\bibitem{Aoki:2006br}
  Y.~Aoki, Z.~Fodor, S.~D.~Katz and K.~K.~Szabo,
  arXiv:hep-lat/0609068.

\end{thebibliography}


\end{document}